\documentclass{pasj00}
\begin{document}
\SetRunningHead
{J.\ Fukue and C.\ Akizuki}
{Relativistic Radiative Flow in a Luminous Disk II}
\Received{yyyy/mm/dd}
\Accepted{yyyy/mm/dd}

\title{Relativistic Radiative Flow in a Luminous Disk II}

\author{Jun \textsc{Fukue} and Chizuru \textsc{Akizuki}} 
\affil{Astronomical Institute, Osaka Kyoiku University, 
Asahigaoka, Kashiwara, Osaka 582-8582}
\email{fukue@cc.osaka-kyoiku.ac.jp}



\maketitle


\begin{abstract}
Radiatively-driven transfer flow perpendicular to a luminous disk
is examined in the relativistic regime of $(v/c)^2$,
taking into account the gravity of the central object.
The flow is assumed to be vertical, and
the gas pressure as well as the magnetic field are ignored.
Using a velocity-dependent variable Eddington factor,
we can solve the rigorous equations of the relativistic radiative flow
accelerated up to the {\it relativistic} speed.
%
For sufficiently luminous cases,
the flow resembles the case without gravity.
For less-luminous or small initial radius cases, however,
the flow velocity decreases due to gravity.
Application to a supercritical accretion disk with mass loss
is briefly discussed.
\end{abstract}

\section{Introduction}

Mass outflow from a luminous disk
is a clue to the formation mechanism of
astrophysical jets and winds in the active objects.
In particular, in a supercritical accretion disk,
the disk local luminosity exceeds the Eddington one,
and mass loss from a disk surface 
driven by radiation pressure would take place
(see Kato et al. 1998 for a review of accretion disks).

So far, radiatively driven outflows from a luminous disk
have been extensively studied by many researchers
(Bisnovatyi-Kogan, Blinnikov 1977; Katz 1980; Icke 1980; Melia, K\"onigl 1989; 
Misra, Melia 1993; Tajima, Fukue 1996, 1998; Watarai, Fukue 1999;
Hirai, Fukue 2001; Fukue et al. 2001; Orihara, Fukue 2003),
and by numerical simulations
(Eggum et al. 1985, 1988; Okuda 2002; Ohsuga et al. 2005; Ohsuga 2006).
In almost all of these studies, however,
the luminous disk was treated as an external radiation source,
and the radiation transfer in the flow was not solved.

Radiation transfer in the disk, on the other hand,
was investigated in relation to the structure
of a static disk atmosphere and
the spectral energy distribution from the disk surface
(e.g., Meyer, Meyer-Hofmeister 1982; Cannizzo, Wheeler 1984;
Shaviv, Wehrse 1986; Adam et al. 1988;
Hubeny 1990; Ross et al. 1992; Artemova et al. 1996;
Hubeny, Hubeny 1997, 1998; Hubeny et al. 2000, 2001;
Davis et al. 2005; Hui et al. 2005).
In these studies, however,
the vertical movement and the mass loss were not considered.

Recently,
mass outflow as well as radiation transfer have been examined
for the first time in the subrelativistic (Fukue 2005a, 2006a) and
fully relativistic cases (Fukue 2005b).
In the latter case,
it is pointed out some singular behavior 
inherent in the relativistic radiative flow
(e.g., Turolla, Nobili 1988; Turolla et al. 1995; Dullemond 1999).
When the gaseous flow is radiatively 
accelerated up to the relativistic regime,
the velocity gradient becomes very large in the direction of the flow.
As a result, 
the radiative diffusion may become {\it anisotropic}
in the comoving frame of the gas.
Hence, 
in a flow that is accelerated from subrelativistic to relativistic regimes,
the Eddington factor should be different from $1/3$
even in the optically thick diffusion limit.

In order to avoid the singular behavior of such a relativistic regime,
for a plane-parallel case, Fukue (2006b) proposed
a {\it velocity-dependent Eddington factor},
which depends on the flow velocity $v$:
\begin{equation}
   f(\beta) = \frac{1+2\beta}{3},
\end{equation}
where $\beta=v/c$.
In Fukue (2006b) this form (1) was adopted
as the simplest one among various forms,
which satisfy several necessary conditions to avoid the singularity.
Physically speaking, this form (1) can be interpreted as follows.
In the high velocity regime, where the radiative diffusion
may become anisotropic in the comoving frame,
the `apparent' optical depth $\tau$ would be of the order of
\begin{equation}
   1+\tau = \frac{1}{\beta}.
\end{equation}
That is, as the flow is accelerated and approaches to the speed of light,
the optical depth becomes zero (outward {\it peaking}).
In this case, the form (1) can be read as
\begin{equation}
   f(\tau) = \frac{3+\tau}{3+3\tau},
\end{equation}
which recovers a similar form of a usual variable Eddington factor
(see, e.g., Tamazawa et al. 1975).
Hence, the applicability and accuracy of the form (1)
 from a low speed regime to a high speed one
would be similar to those of a variable Eddington factor
 from an optically thick regime to an optically thin one.

Similarly, for a spherically symmetric case,
Akizuki and Fukue (2006) proposed a variable Eddington factor,
which depends on the flow velocity $\beta$ 
as well as the optical depth $\tau$:
\begin{equation}
   f(\tau, \beta) = \frac{\gamma(1+\beta)+\tau}{\gamma(1+\beta)+3\tau},
\end{equation}
where $\gamma$ ($=1/\sqrt{1-\beta^2}$) is the Lorentz factor.

In Fukue (2006b),
the plane-parallel case was examined as an example,
although the gravity of the central object was ignored for simplicity.
In this paper,
we thus consider the radiatively driven vertical outflow
-- {\it moving photosphere} -- in a luminous flat disk
within the framework of radiation transfer
in the relativistic regime using $f(\beta)$,
while taking into account the gravity of the central object,
although the gas pressure is ignored.

In the next section
we describe the basic equations in the vertical direction.
In section 3
we show our numerical examination of the radiative flow.
In section 4
we briefly apply the present model
to the case of a supercritial accretion disk.
The final section is devoted to concluding remarks.


\section{Basic Equations and Boundary Conditions}

Let us suppose a luminous flat disk, inside of which
gravitational or nuclear energy is released
via viscous heating or other processes.
The radiation energy is transported in the vertical direction,
and the disk gas, itself, also moves in the vertical direction
due to the action of radiation pressure.
For simplicity, in the present paper,
the radiation field is considered to be sufficiently intense that
the gas pressure can be ignored:
tenuous cold normal plasmas in the super-Eddington disk,
cold pair plasmas in the sub-Eddington disk, or
dusty plasmas in the sub-Eddington disk.
As for the order of the flow velocity $v$,
we consider the fully relativistic regime,
where the terms are retained up to the second order of $(v/c)$.

\subsection{Basic Equations}

Under these assumptions,
the radiation hydrodynamic equations
for steady vertical ($z$) flows without rotation are described as follows
(Kato et al. 1998; Fukue 2006b).

The continuity equation is
\begin{equation}
   \rho cu = J ~(={\rm const.}),
\label{rho1}
\end{equation}
where $\rho$ is the proper gas density, $u$ the vertical four velocity, 
$J$ the mass-loss rate per unit area,
and $c$ the speed of light.
The four velocity $u$ is related to the proper three velocity $v$ by
$u=\gamma v/c$, where $\gamma$ is the Lorentz factor,
$\gamma=\sqrt{1+u^2}=1/\sqrt{1-(v/c)^2}$.

The equation of motion is
\begin{eqnarray}
\!\!\!\!\!
   c^2u\frac{du}{dz} &=& -\frac{GMz}{(R-r_{\rm g})^2 R}
\nonumber
\\
\!\!\!\!\!
         &&       +\frac{\kappa_{\rm abs}+\kappa_{\rm sca}}{c}
                    \left[ F \gamma (1+2u^2) - (cE+cP)\gamma^2 u \right],
\label{u1}
\end{eqnarray}
where $M$ is the mass of the central object,
$R$ $=\sqrt{r^2+z^2}$, $r$ being the radius,
$r_{\rm g}$ ($=2GM/c^2$) is the Schwarzschild radius,
$\kappa_{\rm abs}$ and $\kappa_{\rm sca}$
are the absorption and scattering opacities (gray),
defined in the comoving frame,
$E$ the radiation energy density, $F$ the radiative flux, and
$P$ the radiation pressure observed in the inertial frame.
The first term in the square bracket on the right-hand side
of equation (\ref{u1}) means the radiatively-driven force,
which is modified to the order of $u^2$, whereas
the second term is the radiation drag force,
which is also modified, but roughly proportional to the velocity.
As for the gravity,
we adopt the pseudo-Newtonian potential (Paczy\'nski, Wiita 1980).

When the gas pressure is ignored,
the advection terms of the energy equation 
are dropped (cf. Kato et al. 1998), 
and heating is balanced with the radiative terms,
\begin{equation}
   0 = q^+ -\rho \left( j - \kappa_{\rm abs} cE \gamma^2 
                  - \kappa_{\rm abs} cP u^2
                  + 2 \kappa_{\rm abs} F \gamma u \right),
\label{j1}
\end{equation}
where $q^+$ is the internal heating and $j$ is the emissivity
defined in the comoving frame.
In this equation (\ref{j1}),
the third and fourth terms on the right-hand side
appear in the relativistic regime.

For radiation fields, the zeroth-moment equation becomes
\begin{eqnarray}
   \frac{dF}{dz} &=& \rho \gamma
         \left[ j - \kappa_{\rm abs} cE
                 + \kappa_{\rm sca}(cE+cP)u^2  \right.
\nonumber
\\
    &&   \left. + \kappa_{\rm abs}Fu/\gamma
               -\kappa_{\rm sca}F ( 1+v^2/c^2 )\gamma u \right].
\label{F1}
\end{eqnarray}
The first-moment equation is
\begin{eqnarray}
   \frac{dP}{dz} &=& \frac{\rho \gamma}{c} 
         \left[ \frac{u}{\gamma}j - \kappa_{\rm abs} F
                  + \kappa_{\rm abs}cP \frac{u}{\gamma} \right.
\nonumber
\\
     && \left. -\kappa_{\rm sca}F(1+2u^2)
               +\kappa_{\rm sca}(cE+cP)\gamma u \right].
\label{P1}
\end{eqnarray}

In order to close moment equations for radiation fields,
we need some closure relation.
Instead of the usual Eddington approximation,
we here adopt a {\it velocity-dependent} 
variable Eddington factor $f(\beta)$,
\begin{equation}
   P_0 = f(\beta) E_0
\label{close0}
\end{equation}
in the comoving frame,
where $P_0$ and $E_0$ are the quantities in the comoving frame.
If we adopt this form (\ref{close0}) as the closure relation
in the comoving frame,
the transformed closure relation in the inertial frame is
\begin{equation}
   cP \left( 1 + u^2 - fu^2 \right) = 
   cE \left( f\gamma^2 - u^2 \right) 
   + 2 F \gamma u \left( 1 - f \right),
\label{close}
\end{equation}
or equivalently,
\begin{equation}
   cP \left( 1 - f\beta^2 \right) =
   cE \left( f - \beta^2 \right) + 2F\beta \left( 1 - f \right).
\label{close_beta}
\end{equation}
As a form of the function $f(\beta)$
we adopt the simplest one:
\begin{equation}
   f(\beta) = \frac{1}{3} + \frac{2}{3} \beta
\label{Eddington}
\end{equation}
for a plane-parallel geometry (Fukue 2006b;
cf. Akizuki and Fukue 2006 for a spherically symmetric geometry).

Eliminating $j$ and $E$ 
with the help of equations (\ref{j1}) and (\ref{close}),
equations (\ref{u1}), (\ref{F1}), and (\ref{P1}) become
\begin{eqnarray}
\!\!\!\!\!
\!\!\!\!\!
   c^2u\frac{du}{dz} &=& -\frac{c^2 r_{\rm g}z}{2(R-r_{\rm g})^2 R}
\nonumber
\\
\!\!\!\!\!
\!\!\!\!\!
             &+&    \frac{\kappa_{\rm abs}+\kappa_{\rm sca}}{c} \gamma
                    \frac{F(f\gamma^2+u^2) -cP(1+f)\gamma u}{f\gamma^2-u^2},
\label{u2}
\\
\!\!\!\!\!
\!\!\!\!\!
   \frac{dF}{dz} &=& q^+ \gamma
\nonumber
\\
\!\!\!\!\!
\!\!\!\!\!
    &-&   \rho (\kappa_{\rm abs}+\kappa_{\rm sca}) u
                    \frac{F(f\gamma^2+u^2) -cP(1+f)\gamma u}{f\gamma^2-u^2},
\label{F2}
\\
\!\!\!\!\!
\!\!\!\!\!
   \frac{dP}{dz} &=& q^+ \frac{u}{c}
\nonumber
\\
\!\!\!\!\!
     &-& \rho \frac{\kappa_{\rm abs}+\kappa_{\rm sca}}{c} \gamma
                    \frac{F(f\gamma^2+u^2) -cP(1+f)\gamma u}{f\gamma^2-u^2}.
\label{P2}
\end{eqnarray}

Introducing the optical depth by
\begin{equation}
    d\tau = - ( \kappa_{\rm abs}+\kappa_{\rm sca} ) \rho dz,
\label{tau}
\end{equation}
and using continuity equation (\ref{rho1}),
equations (\ref{u2})--(\ref{tau}) are rearranged as
\begin{eqnarray}
   c^2J\frac{du}{d\tau} &=& \frac{c}{\kappa_{\rm abs}+\kappa_{\rm sca}}
                          \frac{c^2 r_{\rm g}z}{2(R-r_{\rm g})^2 R}
\nonumber
\\
        && -{\gamma}
           \frac{ F(f\gamma^2+u^2) - cP (1+f) \gamma u}
                {f\gamma^2-u^2},
\label{u3}
\\
   J\frac{dz}{d\tau} &=& -\frac{cu}{\kappa_{\rm abs}+\kappa_{\rm sca}},
\label{z3}
\\
   \frac{dF}{d\tau} &=& -\frac{ q^+ }
                              { (\kappa_{\rm abs}+\kappa_{\rm sca}) \rho }
                       \gamma
\nonumber
\\
         && +u
           \frac{ F(f\gamma^2+u^2) - cP (1+f) \gamma u}
                {f\gamma^2-u^2},
\label{F3}
\\
   c\frac{dP}{d\tau} &=&  -\frac{ q^+ }
                { (\kappa_{\rm abs}+\kappa_{\rm sca}) \rho }u
\nonumber
\\
         && +\gamma
           \frac{ F(f\gamma^2+u^2) - cP (1+f) \gamma u}
                {f\gamma^2-u^2}.
\label{P3}
\end{eqnarray}

In this paper 
we assume that the heating takes place deep inside the disk
at the midplane and in the atmosphere $q^+=0$.
However, it is straightfoward to consider more general cases,
where, e.g., the heating $q^+$ is proportional to the gas density $\rho$
(cf. Fukue 2005a, b).

We solved equations (\ref{u3})--(\ref{P3})
for appropriate boundary conditions at the moving surface
with a variable Eddington factor (\ref{Eddington}).

\subsection{Boundary Conditions}

As already pointed out in Fukue (2005a),
the usual boundary conditions for the static atmosphere
cannot be used for the present radiative flow,
which moves with velocity at the order of the speed of light.

When there is no motion in the gas (``static photosphere''),
the radiation field just above the surface
under the plane-parallel approximation is easily obtained.
Namely, just above the disk with surface intensity $I_{\rm s}$,
the radiation energy density $E_{\rm s}$, 
the radiative flux $F_{\rm s}$, and
the radiation pressure $P_{\rm s}$ are
$(2/c)\pi I_{\rm s}$,
$\pi I_{\rm s}$, and 
$(2/3c)\pi I_{\rm s}$, respectively,
where the subscript s denotes the values at the disk surface.
However,
the radiation field just above the surface changes
when the gas itself does move upward (``moving photosphere''),
since the direciton and intensity of radiation
change due to relativistic aberration and Doppler effect
(cf. Kato et al. 1998; Fukue 2000).

Let us suppose a situation that
a flat infinite photosphere with surface intensity $I_{\rm s}$
in the comoving frame is not static,
but moving upward at a speed $v_{\rm s}$ 
($=c\beta_{\rm s}$, and
the corresponding Lorentz factor is $\gamma_{\rm s}$).
Then, just above the surface,
the radiation energy density $E_{\rm s}$, 
the radiative flux $F_{\rm s}$, and
the radiation pressure $P_{\rm s}$ measured in the inertial frame
become, respectively,
\begin{eqnarray}
   cE_{\rm s} 
   &=& {2\pi I_{\rm s}}
       \frac{3\gamma_{\rm s}^2+3\gamma_{\rm s}u_{\rm s}+u_{\rm s}^2}{3},
\label{Es2}
\\
   F_{\rm s}
   &=& {2\pi I_{\rm s}}
       \frac{3\gamma_{\rm s}^2+8\gamma_{\rm s}u_{\rm s}+3u_{\rm s}^2}{6},
\label{Fs2}
\\
   cP_{\rm s}
   &=& {2\pi I_{\rm s}}
       \frac{\gamma_{\rm s}^2+3\gamma_{\rm s}u_{\rm s}+3u_{\rm s}^2}{3},
\label{Ps2}
\end{eqnarray}
where $u_{\rm s}$ ($=\gamma_{\rm s}v_{\rm s}/c$)
is the flow four velocity at the surface (Fukue 2005a).

Thus, we impose the following boundary conditions:
At the flow base (deep ``inside'' the atmosphere)
with an arbitrary optical depth $\tau_0$,
the flow velocity $u$ is zero,
the radiative flux is $F_0$
(which is a measure of the strength of radiation field), and
the radiation pressure is $P_0$
(which connects with the radiation pressure gradient
and relates to the internal structure),
where the subscript 0 denotes the values at the flow base.
At the flow top (``surface'' of the atmosphere)
where the optical depth is zero,
the radiation field should satisfy the values
above a moving photosphere given by
equations (\ref{Es2})--(\ref{Ps2}): i.e.,
\begin{equation}
   \frac{cP_{\rm s}}{F_{\rm s}}
   = \frac{2+6\beta_{\rm s}+6\beta_{\rm s}^2}
       {3+8\beta_{\rm s}+3\beta_{\rm s}^2},
\label{bc2}
\end{equation}
where $\beta_{\rm s}$ is a final speed at the disk surface.

Physically speaking,
in the radiative flow starting from the flow base
with an arbitrary optical depth $\tau_0$,
for initial values of $F_0$ and $P_0$ at the flow base,
the final value of the flow velocity $v_{\rm s}$ at the flow top
can be obtained by solving basic equations.
Furthermore, the mass-loss rate $J$
is determined as an eigenvalue
so as to satisfy the bondary condition (23) at the flow top
(cf. Fukue 2005a in the subrelativistic regime).

However, the permitted region for $J$ is very tight,
and it is difficult to search the value of $J$.
Hence, in this paper, as a mathematically equivalent way,
we fix the value of $J$, and search the value of $P_0$
so as to satisfy the boundary condition (23).

\section{Relativistic Radiative Flow under Gravity}

In this section
we show the relativistic radiative vertical flow in the luminous disk
under the influence of gravity of the central object.
In order to obtain the solution,
as already stated,
we numerically solve equations (\ref{u3})--(\ref{P3}),
starting from $\tau=\tau_0$ at $z=0$
with appropriate initial conditions for $v$, $F$, and $P$,
down to $\tau=0$ so as to satisfy
the boundary conditions (\ref{bc2}) there.
The parameters are
the initial radius $r$ on the disk,
the initial optical depth $\tau_0$,
which relates to the disk surface density,
the initial flux $F_0$,
which is the measure of the strength
of radiation field to gravity,
and the initial radiation pressure $P_0$ at the disk base,
which connects with the radiation pressure gradient
in the vertical direction and
relates to the disk internal structure.
The mass-loss rate $J$ is determined
as an eigenvalue of the boundary condition at the flow top.

Several examples of numerical calculations
are shown in figures 1--3.
Physical quantities are normalized in terms of
the speed of light $c$, the Schwarzschild radius $r_{\rm g}$, and
the Eddington luminosity $L_{\rm E}$
[$=4\pi cGM/(\kappa_{\rm abs}+\kappa_{\rm sca})$].
The units of $F$, $cP$, and $c^2J$ is $L_{\rm E}/(4\pi r_{\rm g}^2)$.
It should be noted that the solutions can be obtained
for arbitrary optical depths $\tau_0$ at the flow base
(see Fukue 2006b, and section 4).
We here show, however, the cases of $\tau_0=1$,
where the velocity change is remarkable.

\begin{figure}
  \begin{center}
  \FigureFile(80mm,80mm){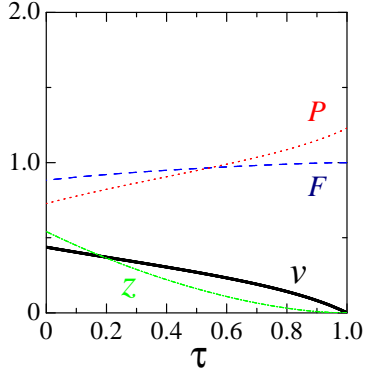}
  \end{center}
\caption{
Flow velocity $v$ (thick solid curve),
flow height $z$ (chain-dotted one), 
radiative flux $F$ (dashed one),
and radiation pressure $P$ (dotted one),
as a function of the optical depth $\tau$.
The parameters are $r=3$, $\tau_0=1$,
$F_0=1$, and $P_0=1.23$.
The mass-loss rate is $J=1$.
The quantities are normalized in units of $c$, $r_{\rm g}$, and 
$L_{\rm E}/(4\pi r_{\rm g}^2)$.
}
\end{figure}

In figure 1
we show the flow velocity $v$ (solid curve),
the flow height $z$ (chain-dotted one), 
the radiative flux $F$ (dashed one),
and the radiation pressure $P$ (dotted one),
as a function of the optical depth $\tau$
for $r=3$, $\tau_0=1$, $F_0=1$, and $P_0=1.23$
(i.e., $J=1$).

As the optical depth decreases from the flow base at the disk equator
to the flow top at the disk surface,
the radiative flux slightly decreases while
the flow velocity increases;
the radiative energy is converted to the flow bulk motion
in the vertical direction.
As usually known, in a static plane-parallel atomosphere,
under the radiative equilibrium with gray approximation,
the vertical flux $F$ is conserved without heating source.
In the present relativistically {\it moving atmosphere}, on the contrary,
the radiative flux $F$ decreases via $Fu$ term,
which acts to drive the gas toward the vertical direction.
In the case of figure 1,
the initial flux ($F_0=1$) is nearly the local Eddington one,
and therefore,
the final flow velocity is mildly relativistic
due to the effect of gravity of the central object.
Other parameter dependences are shown in figures 2 and 3.

\begin{figure}
  \begin{center}
  \FigureFile(80mm,80mm){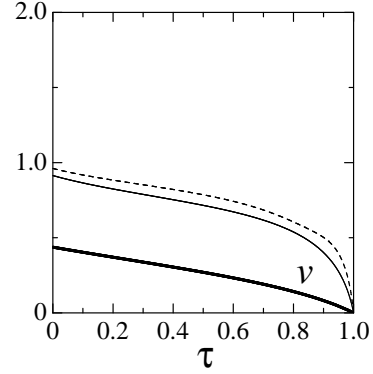}
  \end{center}
\caption{
Flow velocity $v$
as a function of the optical depth $\tau$.
A thick solid curve is for the typical case of
$r=3$, $\tau_0=1$, $F_0=1$, and $P_0=1.23$ (i.e., $J=1$).
A solid curve is for the case of
$r=3$, $\tau_0=1$, $F_0=10$, and $P_0=10.6$ (i.e., $J=1$), while
a dashed one is for the case of
$r=3$, $\tau_0=1$, $F_0=1$, and $P_0=1.041$ (i.e., $J=0.005$).
The quantities are normalized in units of $c$, $r_{\rm g}$, and 
$L_{\rm E}/(4\pi r_{\rm g}^2)$.
}
\end{figure}

In figure 2 the flow velocity $v$ are shown for several parameter set:
A thick solid curve is for the typical case of
$r=3$, $\tau_0=1$, $F_0=1$, and $P_0=1.23$ (i.e., $J=1$).
A solid curve is for the case of
$r=3$, $\tau_0=1$, $F_0=10$, and $P_0=10.6$ (i.e., $J=1$), while
a dashed one is for the case of
$r=3$, $\tau_0=1$, $F_0=1$, and $P_0=1.041$ (i.e., $J=0.005$).

As is easily expected,
the flow velocity remarkably increases
when the initial flux $F_0$ is large
(a solid curve).
When the initial flux is small, on the other hand,
the flow velocity becomes small.

Even for the same initial flux,
when the mass-loss rate is small,
the flow velocity remarkably increases (a dashed curve).
This is because for a small mass-loss rate 
the density decreases, and therefore,
the vertical height $z$ becomes large,
compared with the case for a large mass-loss rate 
with the same optical depth.
As a result, the gas is accelerated along the long distance
to reach the highly relativistic regime.

\begin{figure}
  \begin{center}
  \FigureFile(80mm,80mm){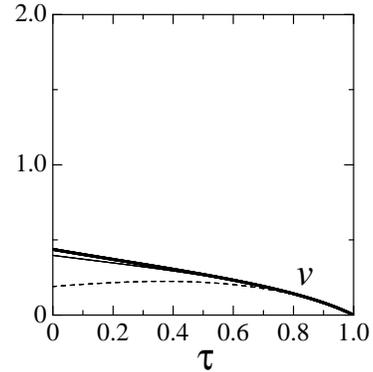}
  \end{center}
\caption{
Flow velocity $v$
as a function of the optical depth $\tau$
for several initial radii $r$:
$r=3$ (a thick solid curve),
$r=2$ (a solid one), and
$r=1.5$ (a dashed one).
Other parameters are
$\tau_0=1$, $F_0=1$, and $P_0=1.23$ (i.e., $J=1$).
The quantities are normalized in units of $c$, $r_{\rm g}$, and 
$L_{\rm E}/(4\pi r_{\rm g}^2)$.
}
\end{figure}

In figure 3
the dependence on the initial radius $r$ is shown:
$r=3$ (a thick solid curve),
$r=2$ (a solid one), and
$r=1.5$ (a dashed one).
Other parameters are
$\tau_0=1$, $F_0=1$, and $P_0=1.23$ (i.e., $J=1$).

For the same initial conditions,
the flow velocity decreases 
as the initial radius decreases.
This is just the effect of gravity of the central object.
In the case of $r=1.5$,
the velocity is slightly {\it decelarated} toward the surface
due to gravity.


\section{Relativistic Radiative Flow in the Critical Disk}

In this section
we apply the present model to
the mass outflow in the luminous supercritical accretion disks
(cf. Fukue 2006a for a subrelativistic case).

When the mass-accretion rate $\dot{M}$ in the disk
around a central object of mass $M$ highly exceeds
the critical rate $\dot{M}_{\rm crit}$,
defined by $\dot{M}_{\rm crit} \equiv L_{\rm E}/c^2$,
the disk is believed to be in the supercritial regime,
and the disk luminosity exceeds the Eddington one.
Such a supercritical accretion disk, a so-called slim disk,
has been extensively studied, both numerically and analytically
(Abramowicz et al. 1988; Eggum et al. 1988;
Szuszkiewicz et al. 1996; Beloborodov 1998;
Watarai, Fukue 1999; Watarai et al. 2000; Mineshige et al. 2000;
Fukue 2000; Kitabatake et al. 2002;
Okuda 2002; Ohsuga et al. 2002, 2003, 2005; 
Watarai, Mineshige 2003; Fukue 2004; Ohsuga 2006).
It was found that
the optically-thick supercritical disk is roughly expressed
by a self-similar model
(e.g., Watarai, Fukue 1999; Fukue 2000; Kitabatake et al. 2002;
Fukue 2004).
Except for the case of Fukue (2004), however,
many of these analytical models
did not consider the mass outflow from the disk surface.
Hence, in this paper
we adopt the model developted by Fukue (2004),
as a background supercritical disk model.

In the {\it critical accretion disk} constructed by Fukue (2004),
the mass-accretion rate is assumed to be regulated 
just at the critical rate with the help of wind mass-loss.
Outside some critical radius,
the disk is in a radiation-pressure dominated standard state,
while inside the critical radius
the disk is in a critical state,
where the excess mass is expelled by wind
and the accretion rate is kept to be just at the critical rate
at any radius.
Here, the critical radius is derived as
\begin{equation}
   r_{\rm cr} = \frac{9\sqrt{3}\sigma_{\rm T}}{16\pi m_{\rm p}c}
                \dot{M}_{\rm input}
              \sim 1.95 \dot{m} r_{\rm g},
\label{rcr}
\end{equation}
where $\dot{M}_{\rm input}$ is the accretion rate
at the outer edge of the disk, and
$\dot{m} = \dot{M}_{\rm input}/\dot{M}_{\rm crit}$.
Outside $r_{\rm cr}$, the accretion rate is constant,
while, inside $r_{\rm cr}$ the accretion rate would vary as
\begin{equation}
   \dot{M}(r) = \frac{16\pi cm_{\rm p}}{9\sqrt{3}\sigma_{\rm T}}r
   =\dot{M}_{\rm input} \frac{r}{r_{\rm cr}}.
\label{dotM}
\end{equation}

In such a critical accretion disk,
the disk thickness $H$ is conical as
\begin{equation}
   \frac{H}{r} = \sqrt{c_3} = \frac{1}{4}
                  \ln \left( 1 + \frac{\dot{m}}{20} \right),
\end{equation}
where $c_3$ is some numerical coefficient
determined by the similar procedure in Narayan and Yi (1994)
for optically-thin advection-dominated disks.
The second equality of this relation
comes from the numerical calculation (Watarai et al. 2000).
Although the mass loss was not considered in Watarai et al. (2000),
we adopted this relation as some measure:
when the normalized accretion rate $\dot{m}$ is 1000,
the coefficient $\sqrt{c_3}$ becomes 0.983.

\begin{figure}
  \begin{center}
  \FigureFile(80mm,80mm){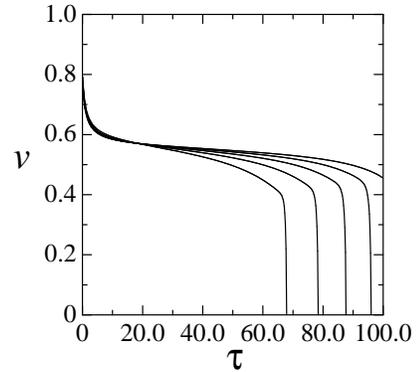}
  \end{center}
\caption{
Flow velocity $v$
as a function of $\tau$
for several values of $r$.
The values of $r$ are, from left to right,
3, 4, 5, 6, and 7.
The quantities are normalized in units of $c$ and $r_{\rm g}$.
The parameters of the critical disk is
$\dot{m}=1000$ and $\alpha=1$.
}
\end{figure}

Furthermore, in Fukue (2004), several alternatives are discussed,
and some of them gives the physical quantities
of the critical accretion disk with mass loss as
\begin{eqnarray}
   \tau_0 &=& \frac{16\sqrt{6}}{\alpha} 
            \sqrt{\frac{r}{r_{\rm g}}},
\\
   F_0    &=& \frac{12\sqrt{6}}{\alpha}\sqrt{c_3}
                 \frac{L_{\rm E}}{4\pi r^2}\sqrt{\frac{r}{r_{\rm g}}},
\\
   cP_0 &=& \frac{cGM}{\kappa}\sqrt{c_3}\tau_0 \frac{1}{r^2} =
                 \frac{16\sqrt{6}cGM}{\alpha \kappa}\sqrt{c_3}
                 \frac{1}{r^2}\sqrt{\frac{r}{r_{\rm g}}},
\end{eqnarray}
where $\alpha$ is the viscous parameter.

In the present non-dimensional unit
in terms of $c$, $r_{\rm g}$, and the Eddington luminosity $L_{\rm E}$,
these physical quantities are expressed as
\begin{eqnarray}
   H &=& \sqrt{c_3} r
\\
   \tau_0 &=& \frac{16\sqrt{6}}{\alpha} 
            \sqrt{{r}},
\\
   F_0 &=& \frac{12\sqrt{6}}{\alpha}\sqrt{c_3}\frac{1}{r^{3/2}},
\\
   P_0 &=& \frac{16\sqrt{6}}{\alpha}\sqrt{c_3} \frac{1}{r^{3/2}},
\end{eqnarray}
where the symbol ``hat'' (say, $\hat{r}$) is dropped.

Using these relations,
we can solve the basic equations, and obtain numerical solutions
at each radius $r$.
The example in the case of $\dot{m}=1000$ and $\alpha=1$
is shown in figures 4 and 5.

In figure 4
we show
the flow velocity $v$
as a function of optical depth $\tau$
for several values of $r$.
The quantities are normalized in units of $c$ and $r_{\rm g}$.
The parameters of the critical disk is
$\dot{m}=1000$ and $\alpha=1$.

As can be seen in figure 4,
the flow velocity $v$ varies self-similarly
for different values of radii.
This may be reflected the initial self-similar models.
As a result,
in each radius with different optical depth,
the flow final speed is almost same.

\begin{figure}
  \begin{center}
  \FigureFile(80mm,80mm){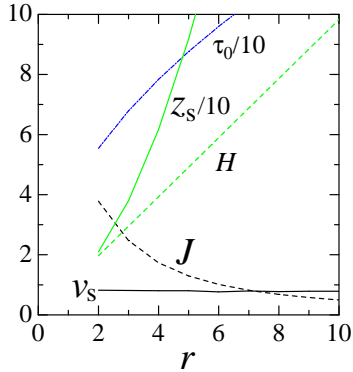}
  \end{center}
\caption{
Several quantities for each radius $r$:
The disk height $H$ (dashed curve) and 
the optical depth $\tau$ (chain-dotted one)
are from the critical model, while
the height $z_{\rm s}$ and velocity $v_{\rm s}$
(solid curves) at the flow top
and the mass-loss rate $J$ (dashed one) are the results of 
the present numerical calculations.
The quantities are normalized in units of $c$ and $r_{\rm g}$.
The parameters of the critical disk is
$\dot{m}=1000$ and $\alpha=1$.
}
\end{figure}

In figure 5
we show several quantities for each radius $r$:
The disk height $H$ (dashed curve) and 
the optical depth $\tau_0$ (chain-dotted one)
are from the critical accretion disk model, while
the height $z_{\rm s}$ and velocity $v_{\rm s}$
(solid curves) at the flow top
and the mass-loss rate $J$ (dashed one) 
are the results of numerical calculations.
The quantities are normalized in units of $c$ and $r_{\rm g}$.
The parameters of the critical disk is
$\dot{m}=1000$ and $\alpha=1$.

As can be seen in figure 5,
the flow height $z_{\rm s}$ is enormously large.
Hence, rigorously speaking,
the present plane-parallel approximation violates
in this application of $\dot{m}=1000$,
and two-dimensional numerical simulation should be needed
in such a case.
However, we can see several insights from the present case.

First,
the final speed of the flow accelerated
in the luminous critical disk
becomes sufficiently relativistic.
In other words, relativistic jets can form
in such a luminous accretion disk.
In addition, this final speed
does not depend on the initial radius so much
due maybe to the initial self-similarity.
Second, on contrary to the final speed,
the mass-loss rate per unit area
increases as the initial radius decreases;
it is roughly approximated by $J \sim 6/r$.
On the other hand, the model mass-loss rate (Fukue 2004, 2006a)
becomes $J=1/r$,
that is qualitatively same, but quantitatively different from
the present numerical values.
Hence, the mass loss from the critical disk
may be concentrated in the inner region,
although the true mass-loss rate cannot be determined
at the present simple state.
This nature, however, is also convenient for
centrally concentrated jets.

\section{Concluding Remarks}

In this paper 
we have examined the relativistic radiative transfer flow in a luminous disk
in the relativistic regime of $(v/c)^2$,
taking account of gravity of the central object.
In such a relativistic regime,
we adopt the velocity-dependent variable Eddington factor.
The flow is assumed to be vertical, and
the gas pressure is ignored for simplicity.
The basic equations are numerically solved
as a function of the optical depth $\tau$,
and the flow velocity $v$, the height $z$, the radiative flux $F$, and
the radiation pressure $P$ are obtained
for a given radius $r$, the initial optical depth $\tau_0$,
and the initial coditions at the flow base (disk ``inside''),
whereas the mass-loss rate $J$ is determined as an eigenvalue
of the boundary condition at the flow top (disk ``surface'').
For sufficiently luminous cases,
the flow resembles the case without gravity.
For less-luminous cases, however,
the flow velocity decreases.

Application to the critical accretion disk was also examined.
If the disk thickness becomes so large,
the present plane-parallel approximation violates
and other treatment, such as numerical simulations, should be needed.

The radiative flow investigaed in the present paper
must be a quite {\it fundamental problem} for
accretion-disk physics and astrophysical jet formation,
although there are many simplifications at the present stage.
For example, we have ignored the gas pressure.
In general cases, where the gas pressure is considered,
there usually appears sonic points,
and the flow is accelerated from subsonic to supersonic.
In this paper we consider a purely vertical flow,
and the cross section of the flow is constant.
If the cross section of the flow increases along the flow,
the flow properties such as a transonic nature would be influenced.
Moreover, we do not consider the rotation of the gas.
In accretion disks around a black hole,
the gas usually rotates around the hole at the relativistic speed.
In the vicinity of the equator,
the vertical flow approximation safely holds,
since the radial gravity is balanced with the centrifugal force.
When the flow is accelerated to be lift up the large height,
the streamline would be curved outward,
since the centrifugal force dominates.
This would violate the vetical flow approximation.

There remain many problems to be solved.

\vspace*{1pc}

The authors thank an anonymous referee for useful comments,
which greatly improved the original manuscript.
This work has been supported in part
by a Grant-in-Aid for the Scientific Research (18540240 J.F.) 
of the Ministry of Education, Culture, Sports, Science and Technology.


\end{document}